\documentclass[10pt, conference]{IEEEtran}
\IEEEoverridecommandlockouts
\usepackage{cite}
\usepackage{amsmath,amssymb,amsfonts}
\usepackage{algorithmic}
\usepackage{graphicx}
\usepackage{textcomp}
\usepackage{xcolor}
\usepackage{listings}
\usepackage{csquotes}
\usepackage[hyphens]{url}
\usepackage[breaklinks=true, colorlinks=true, allcolors=blue, pdfencoding=auto, psdextra, pagebackref, linktoc=all, hyperfootnotes=false, pdfa=true, pdftex]{hyperref}
\usepackage{cleveref}
\usepackage{orcidlink}
\usepackage{pgfplots}
\pgfplotsset{compat=1.18}
\usepackage{tikz}
\usetikzlibrary{shapes.geometric, arrows, fit, positioning, backgrounds}
\tikzstyle{startstop} = [rectangle, rounded corners, text centered, draw=black, fill=none]
\tikzstyle{io} = [trapezium, trapezium left angle=70, trapezium right angle=110, text centered, draw=black, fill=none]
\tikzstyle{process} = [rectangle, text centered, draw=black, fill=none,]
\tikzstyle{decision} = [diamond, align=center, inner sep=0.5pt, draw=black, fill=none]
\tikzstyle{arrow} = [thick,->,>=stealth]
\tikzstyle{dashedbox} = [draw=black, dashed, rounded corners, inner sep=6pt, fill=none]
\tikzstyle{llm} = [rectangle, draw=black, line width=1pt, align=center, fill=black!10]
\tikzstyle{question} = [text centered]
\tikzstyle{categorization} = [circle, draw=black, text centered]
\usepackage{textcomp}
\def\BibTeX{{\rm B\kern-.05em{\sc i\kern-.025em b}\kern-.08em
    T\kern-.1667em\lower.7ex\hbox{E}\kern-.125emX}}

\ifdefined\backref
\renewcommand*{\backref}[1]{}% for backref < 1.33 necessary
\renewcommand*{\backrefalt}[4]{%
	\ifcase #1 (Not cited.)% no citations
	\or        (Cited on page:~#2)% one citation
	\else      (Cited on pages:~#2)% many citations
	\fi%
}
\else
\fi
\begin{document}

\title{Using Large Language Models to Support Automation of Failure Management in CI/CD Pipelines: A Case Study in SAP HANA}

\author{\IEEEauthorblockN{Duong Bui \orcidlink{0009-0006-1605-3264}$^{*}$}
\IEEEauthorblockA{ \textit{SAP}\\
duong.bui@sap.com\\
}
\and
\IEEEauthorblockN{Stefan Grintz \orcidlink{0009-0002-8432-3650}$^{*}$}
\IEEEauthorblockA{ \textit{SAP}\\
stefan.grintz@sap.com\\
}
\and
\IEEEauthorblockN{Alexander Berndt \orcidlink{0009-0009-5248-6405}}
\IEEEauthorblockA{\textit{Heidelberg University}\\
alexander.berndt@uni-heidelberg.de\\
}
\and
\IEEEauthorblockN{Thomas Bach \orcidlink{0000-0002-9993-2814}}
\IEEEauthorblockA{\textit{SAP}\\
thomas.bach03@sap.com \\
\thanks{$^*$Both authors contributed equally to this work.}
}}

\maketitle

\begin{abstract}
    CI/CD pipeline failure management is time-consuming when performed manually. Automating this process is non-trivial because the information required for effective failure management is unstructured and cannot be automatically processed by traditional programs. With their ability to process unstructured data, large language models (LLMs) have shown promising results for automated failure management by previous work. Following these studies, we evaluated whether an LLM-based system could automate failure management in a CI/CD pipeline in the context of a large industrial software project, namely SAP HANA. We evaluated the ability of the LLM-based system to identify the error location and to propose exact solutions that contain no unnecessary actions. To support the LLM in generating exact solutions, we provided it with different types of domain knowledge, including pipeline information, failure management instructions, and data from historical failures. We conducted an ablation study to determine which type of domain knowledge contributed most to solution accuracy. The results show that data from historical failures contributed the most to the system’s accuracy, enabling it to produce exact solutions in 92.1\% of cases in our dataset. The system correctly identified the error location with 97.4\% accuracy when provided with domain knowledge, compared to 84.2\% accuracy without it. In conclusion, our findings indicate that LLMs, when provided with data from historical failures, represent a promising approach for automating CI/CD pipeline failure management.
\end{abstract}

\begin{IEEEkeywords}
CI/CD, failure remediation, automation, large language models, case study
\end{IEEEkeywords}

\section{Introduction}
% Length: Limit to the 1st page
CI/CD pipelines are widely used in modern software development because they enable the rapid delivery of software by automating many stages of the software development lifecycle~\cite{myllynen_2024}. However, managing CI/CD pipeline failures remains a costly challenge. CI/CD pipelines are dynamic and frequently updated~\cite{myllynen_2024}. Therefore, pipeline resilience patterns, such as retries and circuit breakers~\cite{Gupta_2024}, cannot automatically detect and respond to all types of failures. In many cases, on-call engineers, who are responsible for monitoring and responding to the CI/CD pipeline, must manually identify root causes and resolve failures~\cite{de_2024_automateLogAnalysis, chen_2024, wang_2024_LLMforRCAandSolution, zhang_2024, ahmed_2023}.
% TODO: search the related works to add more citations to the sentence above.

Manually managing failure in CI/CD pipeline is time-consuming~\cite{de_2024_automateLogAnalysis}. It often requires engineers with several years experience who deeply understand the CI/CD pipeline and its behavior to manually inspect lengthy logs filled with irrelevant information~\cite{xu_2025, roy_2024, ahmed_2023, chen_2024}. It is also error-prone, as critical details can easily be overlooked when scanning through such extensive logs~\cite{myllynen_2024}. These challenges can slow down development, waste valuable resources, and cause frustration among both developers and stakeholders~\cite{myllynen_2024, xu_2025}.

Failure-related information, such as logs, is typically unstructured~\cite{chen_2024, xu_2025, ahmed_2023}. This makes automating failure management without LLMs challenging~\cite{xu_2025}. The most prominent approaches rely on machine learning models that require training on large, generalizable datasets to process such unstructured logs~\cite{xu_2025, le_2021_nonLLM, le_2022_nonLLM, du_2017_nonLLM, yang_2021_nonLLM}. However, the accuracy of these approaches is significantly reduced in real-world scenarios when they encounter unseen log formats or when training datasets are small~\cite{liu_2024_logAnalysis}. Real-world CI/CD pipelines are frequently updated to add new features and fix bugs, and updated logs can be incompatible with the data on which the models were trained~\cite{liu_2024_logAnalysis}. Moreover, when a log format is updated, the number of updated samples is small, which makes retraining the models challenging and expensive~\cite{liu_2024_logAnalysis}. Given these limitations, large language models (LLMs), with their ability to process unstructured data without task-specific training, present a promising approach for automating failure-management tasks in CI/CD pipelines.

In this study, we examined whether an LLM could support automating failure management in a CI/CD pipeline for the delivery of SAP HANA. Specifically, this study addresses the following research questions (RQs):
\begin{enumerate}
    \item RQ1: What are the most common causes of failure in the CI/CD pipeline for the delivery of SAP HANA?
    \item RQ2: How accurately can an LLM identify the error location in the CI/CD pipeline for the delivery of SAP HANA?
    \item RQ3: Which type of knowledge contributes most to an LLM’s accuracy in proposing solutions for failures in the CI/CD pipeline for the delivery of SAP HANA?
\end{enumerate}

The structure of this paper is as follows. \Cref{sec_background} introduces the necessary background information for our study, including related work. \Cref{sec_methodology} explains the methodology used to conduct the experiments and answer the research questions. \Cref{sec_experimentResult} presents the experimental results and the corresponding answers to the research questions. \Cref{sec_lessonLearned} discusses the lessons learned from the experimental findings. \Cref{sec_threadsToValidy} describes the potential threats to the validity of our results. Finally, \Cref{sec_conclusion} summarizes our study.

% ----------------------------------------------------------------------------
% Section
% ----------------------------------------------------------------------------
\section{Background} \label{sec_background}
\subsection{SAP HANA}
SAP HANA is an in-memory database management system, developed by SAP for business applications~\cite{faerber_2012}. There are two types of SAP HANA. One is HANA On-Premise, which is installed and runs on SAP customers’ own servers and infrastructure. The other is HANA Cloud, which SAP deploys on a cloud infrastructure to offer it as a service. For HANA On-Premise, new versions are released monthly to provide new features and patches. For HANA Cloud, new releases are produced every week. The delivery of the software is based on a CI/CD process, which consists of several steps, such as building the source code, executing tests, or, in the case of HANA Cloud, creating containers and their deployment on the cloud infrastructure. The delivery process for SAP HANA is implemented via a CI/CD pipeline in Jenkins.

\subsection{Jenkins Pipeline}
Jenkins is an open-source automation tool that allows CI/CD processes to be run through pipelines. A Jenkins pipeline consists of a set of stages, each of which defines a distinct section of the pipeline workflow. Both sequential and parallel arrangements of stages are possible in a Jenkins pipeline. A stage is divided into one or multiple related steps, with each executing a specific task. A step carries out an actual operation, such as a single command call to start a specific test or the execution of a script that processes the entire task for the step. For example, a stage that performs the testing of the developed software can be composed of individual steps such as the installation, the test execution, the test evaluation, the uninstallation, and finally the cleanup.

\begin{lstlisting}[caption={Exemplary Jenkinsfile with stages and steps (adapted from\\\hspace{\textwidth} the Jenkins User Handbook~\cite{jenkins_1})},label={lst:jenkinsfile_1}, basicstyle=\footnotesize\ttfamily]
pipeline {
    agent any
    stages {
        stage('Build') {
            steps {
                sh 'make'
            }
        }
        stage('Test'){
            steps {
                sh 'make check'
            }
        }
        stage('Deploy') {
            steps {
                sh 'make publish'
            }
        }
    }
}
\end{lstlisting}

The stages and steps of a Jenkins pipeline are defined by a domain-specific language (DSL) and stored as Groovy code in a so-called \emph{Jenkinsfile}~\cite{jenkins_1}. \Cref{lst:jenkinsfile_1} shows a simplified example of a Jenkinsfile for a pipeline with stages for build, test, and deploy, and their respective steps.

A Jenkins pipeline can also be further modularized by steps that invoke other pipelines, thus creating a hierarchy of pipelines. A pipeline that is triggered by another one is referred to as \emph{downstream pipeline} or \emph{sub-pipeline} and contains a set of stages with steps defined in its own Jenkinsfile. In the pipeline example at \Cref{lst:jenkinsfile_1}, there are stages for Build, Test, and Deploy, with the required executions specified directly in their respective steps. These steps can also be delegated to another pipeline by integrating the call to dedicated sub-pipelines with the corresponding steps for each stage. In \Cref{lst:jenkinsfile_2}, for example, the steps for the Build, Test, and Deploy stages are each outsourced to corresponding sub-pipelines.

\begin{lstlisting}[caption={Exemplary Jenkinsfile with sub-pipelines},label={lst:jenkinsfile_2}, basicstyle=\footnotesize\ttfamily]
pipeline {
    agent any
    stages {
        stage('Build') {
            steps {
                build job: 'BuildPipeline'
            }
        }
        stage('Test') {
            steps {
                build job: 'TestPipeline'
            }
        }
        stage('Deploy') {
            steps {
                build job: 'DeployPipeline'
            }
        }
    }
}
\end{lstlisting}

\subsection{Jenkins Pipeline Used for the Delivery of SAP HANA}\label{subsec_pipeline_hana}
Accordingly, the Jenkins pipeline for the CI/CD workflow of the SAP HANA delivery process is made up of multiple stages. Each stage either executes a single step directly or triggers a particular sub-pipeline with its own set of stages, each of which also contains a single step. In case of a new HANA Cloud release, there is a main pipeline with $46$ steps, three of them are calls to sub-pipelines. The sub-pipelines serve the purpose of performing special tasks for the container creation for HANA Cloud, which were separated from the main pipeline due to improved modularity and maintainability of the Jenkinsfile code. Of these three sub-pipelines, two have $5$ steps and one has $8$ steps, resulting in a total of $64$ pipeline steps to process the delivery of a HANA Cloud release.

Steps that are executed directly for a specific task - those that do not call a sub-pipeline - are encapsulated in so-called Jenkins \emph{freestyle jobs}. A Jenkins freestyle job, is a single unit for task execution and can be created and configured via the Jenkins web user interface~\cite{jenkins_1}. Within a freestyle job, the actions for the corresponding step, such as calling a command or running a script, are defined together with parameters received from the pipeline and to be passed for execution. These jobs, which are triggered by the pipeline as a step, are hereinafter also referred to as \emph{downstream jobs}.

Integrating such jobs as pipeline steps provides the advantage of a more structured overview of the pipeline's logging. Instead of logging all steps in a single console output of the pipeline, we receive a separate console output for each downstream job and thus for each individual step. In the Jenkins web user interface, the console output of a pipeline provides a link to each completed and the currently running downstream job, which redirects to its console output.

The downstream jobs of a CI/CD pipeline for SAP HANA execute Python scripts that, in most cases, perform API calls to remote services like version control systems, such as Git~\cite{git_1}, code review tools like Gerrit~\cite{gerrit_1}, build servers, databases storing build information, or other resources like issue tracking systems. In addition, some pipeline steps trigger so-called \emph{remote pipelines} on different Jenkins servers. These remote Jenkins servers are dedicated to tasks that are outside the technical scope of the Jenkins server hosting the main pipeline and its sub-pipelines for the SAP HANA delivery process. Consequently, these remote pipelines are also considered as remote services. The team that operates and maintains the pipeline for the SAP HANA delivery has no control over these remote services and resources, as they are managed by various other teams.

\begin{figure}[htbp]
\centering
\includegraphics[width=\columnwidth]{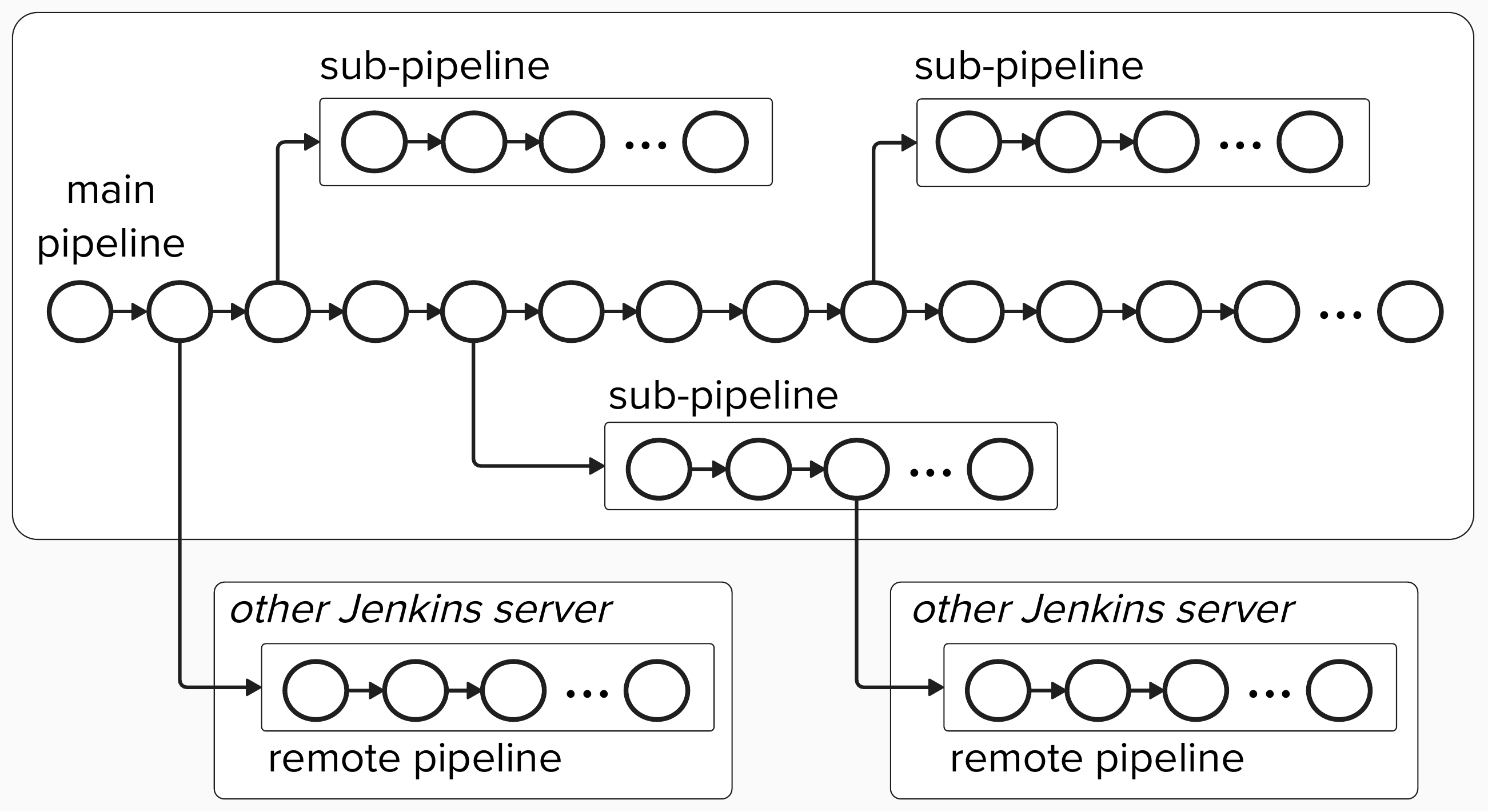}
\caption{Simplified structure of a complex CI/CD pipeline hierarchy}
\label{fig_jenkins_pipeline_1}
\end{figure}

Any error in a downstream job, and therefore in a step, whether in the main pipeline, one of the sub-pipelines, or a remote pipeline, results in an immediate failure and thus aborts the entire pipeline process. The main pipeline can only be restarted after the cause of the error has been resolved. A recovery file contains all the data required to resume the pipeline run, primarily the step at which the pipeline was aborted. Additionally, parameter values required to continue a new pipeline run are maintained there. If a failed pipeline is restarted, all already succeeded steps are skipped, and the new pipeline run continues with repeating the step that failed in the previous run.

In the event of a pipeline failure, the Jenkins web user interface is accessed for error diagnosis, if done manually. First, the console output of the main pipeline is checked to see if there is an error directly there, in one of its sub-pipelines or in a remote pipeline. In all these cases, the pipeline console output contains a line with an error message and a link to the relevant console output. Manual troubleshooting, therefore, requires reading several nested Jenkins console outputs to finally identify the step with the downstream job containing the error message about the cause of the pipeline failure, which is referred to as the \emph{most downstream failed job}.

For Jenkins, there are options to visualize pipelines on the Jenkins web user interface, like \emph{Blue Ocean} or equivalent plugins, which can be helpful for the manual failure diagnosis~\cite{jenkins_2}. However, the benefit of these available features is limited for us because they do not cover our complete CI/CD workflow. For example, they show the steps of the main pipeline but not our entire pipeline hierarchy, including its sub-pipelines and remote pipelines with their steps.

\begin{figure}[htbp]
\centering
\includegraphics[width=\columnwidth]{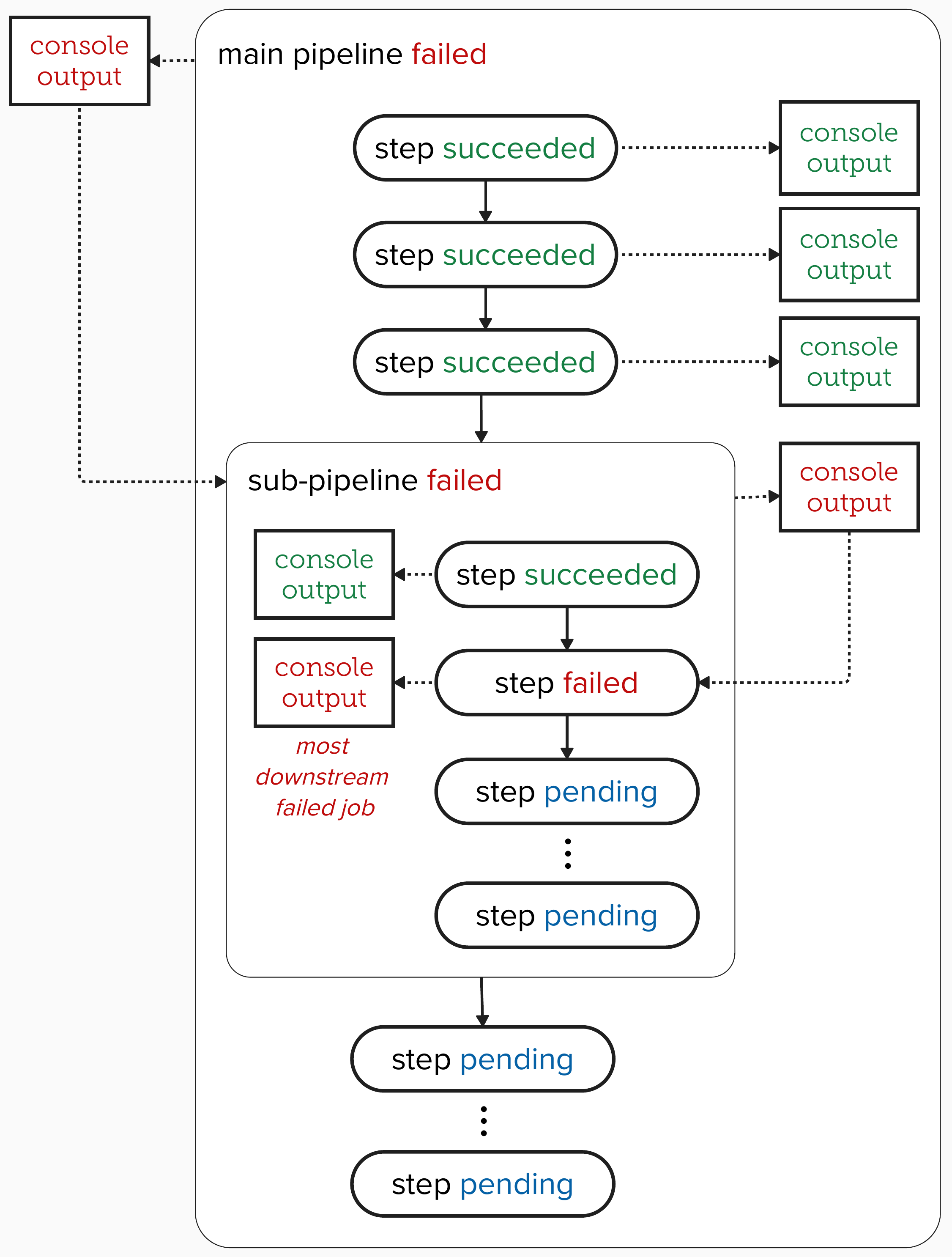}
\caption{Pipeline failure due to a failed step in a sub-pipeline}
\label{fig_jenkins_pipeline_2}
\end{figure}

For automated access to the Jenkins pipeline, we use the Python Jenkins API~\cite{jenkins_api_1}, which provides the functionality to return console outputs but not directly any information about a failed step. Instead, a console output must be parsed by a custom implementation to extract the error information.

% -------------------- Subsection ------------------------------
\subsection{Related Work}
To manage failures in CI/CD pipelines, an essential task is to extract failure-related information from console logs, which record system state and important events. Therefore, many studies investigate the use of LLMs for log analysis. For example, LLMs are used for log parsing - extracting structural templates from logs~\cite{xiao_2024_LogParsing, liu_2024_logAnalysis, ju_2024_logParsing} - and for log anomaly detection to enable early alerts~\cite{egersdoerfer_2023_anomalyDetection, qi_2023_anomalyDetection, liu_2024_logAnalysis, shan_2024_anomalyDetection}. These studies report promising results, demonstrating the capability of LLMs for log analysis. However, they cannot resolve failures directly. They only extract failure-related information from logs to support on-call engineers in troubleshooting. Finding root causes and solutions typically requires software engineers with several years of experience and a deep understanding of pipeline behavior~\cite{xu_2025, roy_2024, ahmed_2023, chen_2024}. Therefore, these tasks demand significant domain-specific knowledge that goes beyond what logs alone can provide.

Most studies on using LLMs for finding root causes focus on cloud-incident domains. Although each study has unique aspects, they share commonalities in how they use LLMs. First, LLMs are used to summarize failure information from different sources, either by supplying the information directly to LLMs~\cite{chen_2024, zhang_2024} or by equipping the LLMs with function-calling capability~\cite{wang_2024} so they can retrieve information themselves~\cite{roy_2024, wang_2024_LLMforRCAandSolution}. Second, these studies provide the LLM with the generated summaries and domain-specific knowledge (e.g., related historical incidents) retrieved via vector search~\cite{Salton_1975_vectorsearch} and ask the LLM to identify root causes~\cite{chen_2024, zhang_2024, roy_2024, wang_2024_LLMforRCAandSolution}.

Regarding failure management in CI/CD pipelines, Chaudhary et al. propose an LLM-based chatbot to answer questions about CI/CD pipelines, including instructions for resolving failures~\cite{chaudhary_2024_LLMChatbot}. Xu et al. propose an end-to-end LLM-based failure-management system, which is the study most similar to our work~\cite{xu_2025}. However, their approach splits failure management into separate steps - root-cause analysis and solution generation - and calls the LLM for each step~\cite{xu_2025}. We argue that this is time and cost inefficient because these steps depend on one another and must be performed sequentially. Moreover, it unnecessarily increases the risk of inaccuracy. If the LLM produces incorrect insights in an earlier step, subsequent steps will also be affected. Separating LLM calls may be reasonable for cloud incidents, where extensive information is needed to determine root causes and solutions. In the case of CI/CD pipelines, however, we argue that finding the root cause and a solution can be accomplished in a single LLM call.

\section{Methodology} \label{sec_methodology}
\begin{figure}[htbp]
\centering
\begin{tikzpicture}[node distance=0.75cm, font=\small]
\node (start) [startstop] {A build failed};
\node (pro1) [process, below of=start] {Retrieve most downstream failed job's console log};
\node (pro2) [process, below of=pro1] {Preprocess log};
\node (dec1) [decision, below of=pro2, yshift=-1cm, text width=3.5cm, aspect=3] {Is historical failures domain knowledge provided?};
\node (pro3a) [process, below of=dec1, xshift=-2cm, yshift=-1.25cm, text width=3.5cm] {Retrieve historical failures with same most downstream failed job};
\node (pro3b) [process, below of=dec1, xshift=2cm, yshift=-2cm, text width=3.5cm] {Get pipeline information/ failure-management instruction knowledge from Knowledge utility class};
\node (pro4) [process, below of=pro3a, yshift=-1cm, text width=3.5cm] {Use RAG to rank top 3 historical failures with most similar console logs};

% Draw the dashed box
\begin{scope}[on background layer]
\node[dashedbox, fit=(dec1) (pro3b) (pro4)] (group) {};
\node[rotate=-90, anchor=center, inner sep=2pt]
at ([xshift=10pt]group.east) {Prepare domain knowledge};
\end{scope}

\node (pro5) [process, below=2mm of group] {Call LLM with log + domain knowledge};
\node (llm1) [llm, below of=pro5] {LLM};
\node (pro7) [process, below of=llm1] {Return root cause and proposed solution};
\node (stop) [startstop, below of=pro7, yshift=-0.25cm, text width=5cm] {Send root cause and proposed solution to on-call engineers};
\draw [arrow] (start) -- (pro1);
\draw [arrow] (pro1) -- (pro2);
\draw [arrow] (pro2) -- (dec1);
\draw [arrow] (dec1) -- node[anchor=east] {yes} (pro3a);
\draw [arrow] (dec1) -- node[anchor=west] {no} (pro3b);
\draw [arrow] (pro3a) -- (pro4);
\draw [arrow] (pro4) -- (pro5);
\draw [arrow] (pro3b) -- (pro5);
\draw [arrow] (pro5) -- (llm1);
\draw [arrow] (llm1) -- (pro7);
\draw [arrow] (pro7) -- (stop);

\end{tikzpicture}
\caption{Control-flow diagram of the LLM-based failure-management system.}
\label{fig_failureManagementSystem}
\end{figure}
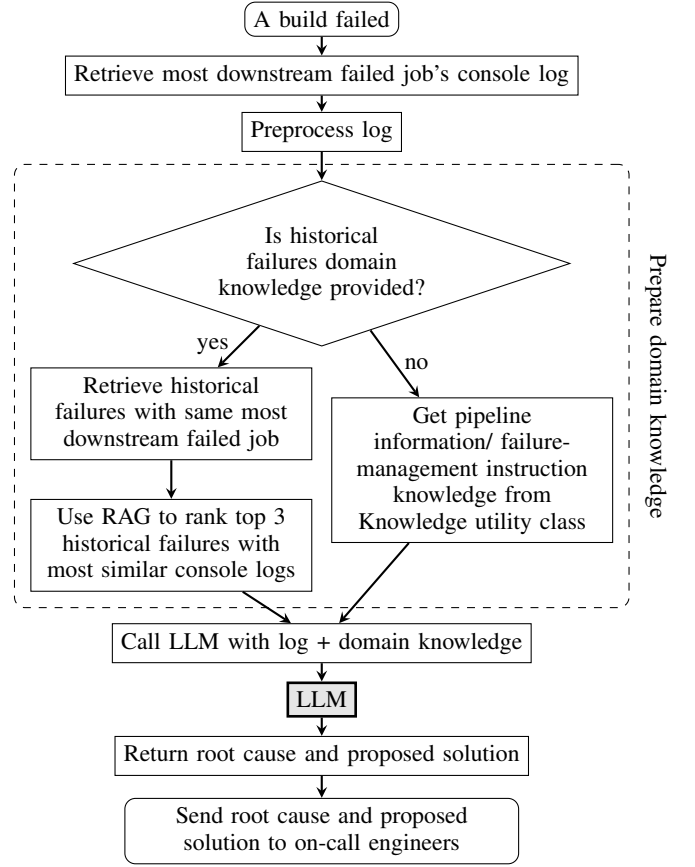

Because we use Jenkins to operate the CI/CD pipeline for the delivery of SAP HANA, we implement the LLM-based system for managing pipeline failures (henceforth referred to as the \emph{failure-management system}) also as a Jenkins pipeline. This allows the failure-management system to be triggered automatically when a CI/CD build fails. We use OpenAI’s GPT-4o model with temperature $0.0$~\cite{gpt4o}. The general tasks of the failure-management system are to automatically find the console logs that contain the root cause (the console logs of the most downstream failed job), preprocess those logs, provide the preprocessed logs and related domain knowledge to the LLM, and ask it to identify the causes of failures and suggest solutions. After receiving the LLM’s answer, the system sends the response to on-call engineers to help them resolve the problem as quickly as possible. \Cref{fig_failureManagementSystem} shows the control-flow diagram of our system. In this section, we provide a detailed description of how our failure-management system operates and outlines the design of our experiment to address the three research questions.

% -------------------- Subsection ------------------------------

\subsection{Most Downstream Failed Job Finding} \label{subsec_mostDownstreamJobFinding}
% Because any changes to the CI/CD pipeline for the delivery of SAP HANA must pass rigorous quality tests, failures are typically caused by remote services or by connection issues to those services, which cannot be tested before execution. In these cases, the CI/CD pipeline console logs indicate only which downstream sub-pipeline, remote pipeline, or job failed, without providing information about the root cause. This information is typically found only in the console log of the most downstream failed job, which triggers the failure in the main pipeline.

CI/CD pipeline console logs show only which downstream sub-pipeline, remote pipeline, or job failed. They do not provide information about the root cause. That information is typically found only in the console log of the most downstream failed job - the job that triggered the failure in the main pipeline. Therefore, to determine the root cause, we must locate the most downstream failed job and analyze its console log.

As described in \Cref{subsec_pipeline_hana}, none of the available tools can directly extract the most downstream failed job, and extracting it from the console logs is the most direct automated approach. Because we observe that the format of the lines that show downstream job information has remained unchanged for a long time, we use regular expressions~\cite{python_re} to search for those lines and extract downstream job details such as pipeline name, build number, and access links.

We start by finding the failed downstream job or sub-pipeline of the main pipeline (denoted DS-1). Then we find the downstream job, sub-pipeline, or remote pipeline of DS-1 and continue until we find the most downstream failed job.

% To reduce the manual effort involved in error handling as described in \Cref{subsec_pipeline_hana}, automating the process of identifying the failure-causing downstream job is essential. For this purpose, an implementation was developed that parses the pipeline console outputs using regular expressions~\cite{python_re}. Starting with the main pipeline, it checks which downstream job or sub-pipeline caused the main pipeline's failure. For example, if the error is located in a downstream job of a sub-pipeline, a log pattern pointing to the failed sub-pipeline would be found in the main pipeline's console output. Then the sub-pipeline's console output would be examined, and finally, the downstream job that led to the main pipeline's failure would be identified. Although this is a straightforward approach to identifying the failure-causing downstream job, we have experienced that changes and updates to the pipeline environment entail frequent adjustments to this implementation, as log patterns in the console outputs also change over time.

% To minimize the maintenance effort of this code, we wanted to find out if an LLM-based approach is an appropriate alternative to automatically identify the failure-causing downstream job.

% -------------------- Subsection ------------------------------

\subsection{Log Preprocessing} \label{subsec_logProcessing}

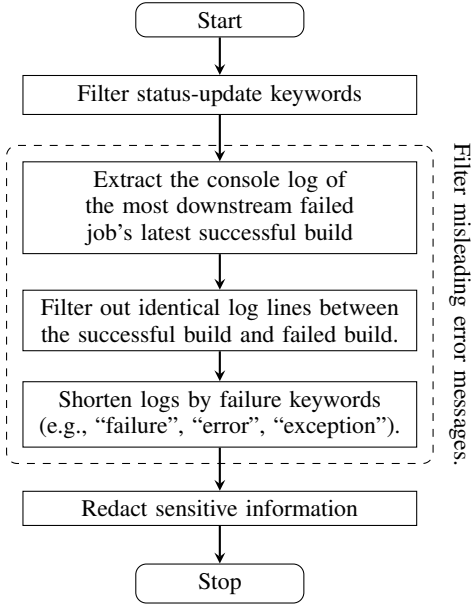
\begin{figure}[htbp]
\centering
\begin{tikzpicture}[node distance=1cm, font=\small]
\node (start) [startstop, text width=2cm] {Start};
\node (pro1) [process, below of=start, text width=5cm] {Filter status-update keywords};
\node (pro2) [process, below of=pro1, text width=5cm, yshift=-0.5cm] {Extract the console log of the most downstream failed job’s latest successful build};
\node (pro3) [process, below of=pro2, text width=5cm, yshift=-0.5cm] {Filter out identical log lines between the successful build and failed build.};
\node (pro4) [process, below of=pro3, text width=5cm, yshift=-0.25cm] {Shorten logs by failure keywords (e.g., “failure”, “error”, “exception”).};
\node (pro5) [process, below of=pro4, text width=5cm, yshift=-0.25cm] {Redact sensitive information};
\node (stop) [startstop, below of=pro5, text width=2cm] {Stop};

% Draw the dashed box
\begin{scope}[on background layer]
\node[dashedbox, fit=(pro2) (pro4)] (group) {};
\node[rotate=-90, anchor=center, inner sep=2pt, fill=white]
at ([xshift=10pt]group.east) {Filter misleading error messages.};
\end{scope}

\draw [arrow] (start) -- (pro1);
\draw [arrow] (pro1) -- (pro2);
\draw [arrow] (pro2) -- (pro3);
\draw [arrow] (pro3) -- (pro4);
\draw [arrow] (pro4) -- (pro5);
\draw [arrow] (pro5) -- (stop);
\end{tikzpicture}
\caption{Control-flow diagram of log preprocessing.}
\label{fig_logPreprocessing}
\end{figure}

An overview of the log processing steps is shown in \Cref{fig_logPreprocessing}. In the following, we explain each step in more detail. To limit token usage, we attempt to shorten the logs we provide to the LLM as much as possible. First, we use regular expressions~\cite{python_re} to filter out Jenkins' status-update keywords (e.g., timestamps, \enquote{echo}). Each of these keywords takes the place of one log line, without any additional information, that helps to identify failure causes.

We follow Xu et al.'s approach of comparing failed logs with successful logs and filtering out identical lines. These lines may contain error or warning messages that appear in every run~\cite{xu_2025}. They do not provide useful information about failures and may even be misleading~\cite{xu_2025}. However, unlike Xu et al., we do not extract log templates from successful builds and store them for later reuse. The CI/CD pipeline for delivering SAP HANA is frequently updated, and its log templates often change. Therefore, extracting and maintaining templates in storage would be costly in our case. We extract the console logs of the latest successful build, filter out unrelated status-update keywords, identify lines in the failed logs that differ from the successful logs by fewer than two words, and remove these lines. We acknowledge that this approach is less accurate than Xu et al.'s approach. However, compared with the large logs of the CI/CD pipeline that Xu et al. process, the logs of the most downstream failed jobs we process are shorter. Therefore, we prefer a cost-efficient approach rather than attempting to comprehensively remove all identical log lines between failed and successful builds.

We then further shorten the console logs by using regular expressions~\cite{python_re} to search for lines containing failure keywords (e.g., \enquote{failure}, \enquote{error}, \enquote{exception}) and retain only each such line, along with five lines before and five lines after it. These surrounding lines provide contextual information for identifying the cause of the failure~\cite{xu_2025}. Finally, we use StarPII~\cite{starpii}, a model trained to detect personally identifiable information, to identify and replace sensitive information in console logs with tags such as NAME, ID, and EMAIL.

% -------------------- Subsection ------------------------------

\subsection{LLM-Based Cause-of-Failure and Solution Finding} \label{subsec_method_causeAndSolutionFinding}

\begin{figure}[htbp]
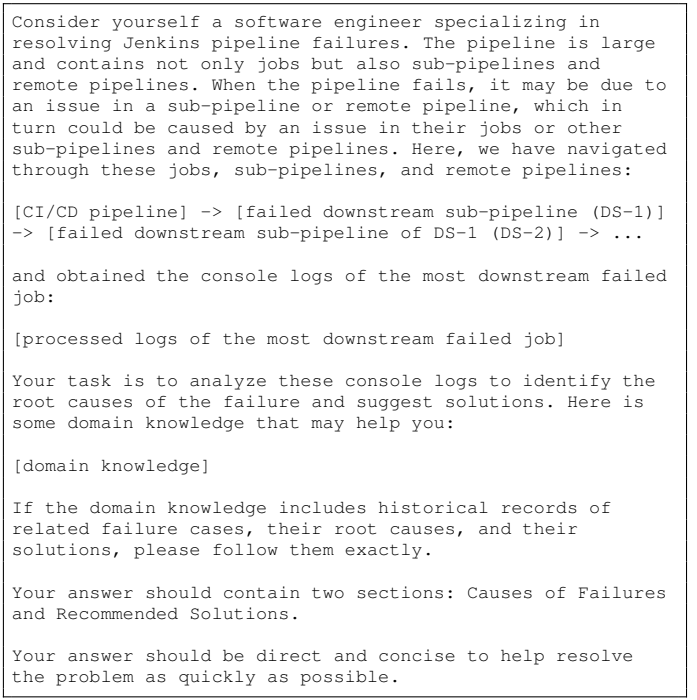

\center
\begin{lstlisting}
Consider yourself a software engineer specializing in resolving Jenkins pipeline failures. The pipeline is large and contains not only jobs but also sub-pipelines and remote pipelines. When the pipeline fails, it may be due to an issue in a sub-pipeline or remote pipeline, which in turn could be caused by an issue in their jobs or other sub-pipelines and remote pipelines. Here, we have navigated through these jobs, sub-pipelines, and remote pipelines:

[CI/CD pipeline] -> [failed downstream sub-pipeline (DS-1)] -> [failed downstream sub-pipeline of DS-1 (DS-2)] -> ...

and obtained the console logs of the most downstream failed job:

[processed logs of the most downstream failed job]

Your task is to analyze these console logs to identify the root causes of the failure and suggest solutions. Here is some domain knowledge that may help you:

[domain knowledge]

If the domain knowledge includes historical records of related failure cases, their root causes, and their solutions, please follow them exactly.

Your answer should contain two sections: Causes of Failures and Recommended Solutions.

Your answer should be direct and concise to help resolve the problem as quickly as possible.
\end{lstlisting}
\caption{Prompt template for finding cause and solution of a failure}
\label{promptTemplate_rq3}
\end{figure}

In the prompt to the LLM, we use a role-prompting technique~\cite{Schulhoff_2025_rolePrompting} to have the LLM act as a software engineer who aims to find the causes of failures and their solutions as soon as possible. We provide related information about our pipeline, the failed build structure, and the console logs of the most downstream failed job, which were processed using the methods mentioned in \Cref{subsec_logProcessing}, along with related domain knowledge. The prompt template we used is shown in \Cref{promptTemplate_rq3}.

\lstset{
    language={}, % plain text (or Python if you want highlighting)
    basicstyle=\ttfamily\scriptsize, % smaller monospaced font
    columns=fullflexible, % better monospaced spacing
    keepspaces=true, % preserve multiple spaces
    upquote=true,
    breaklines=true, % enable wrapping
    breakatwhitespace=true, % prefer breaking at spaces
    breakindent=0pt, % no hanging indent for wrapped lines
    xleftmargin=0pt, % flush with the left page margin
    xrightmargin=0pt,
    frame=single,
    stepnumber=1,
    tabsize=4,
    % aboveskip=2pt,   % space above the listing
    belowskip=0pt    % space below the listing
}

We provide three types of domain knowledge: pipeline information, failure-management instructions, and historical records of related failed builds.

\textbf{Pipeline information (PI)}: This includes information about our CI/CD pipeline that could help identify root causes of failures and suggest solutions, such as pipeline structure, naming conventions, sub-pipeline names, remote pipeline URLs, and the effects of sub-pipeline or remote pipeline failures. We store this domain knowledge as a class-level attribute in a Knowledge utility class dedicated to this purpose, and access these attributes whenever we provide the LLM with domain knowledge. This enables easy access and reuse. Moreover, because all knowledge is centralized in a single utility class, we need only update that class for changes to take effect everywhere the knowledge is used.

\textbf{Failure-management instructions (FMI)}: This includes the steps an experienced on-call engineer would take to manage a failed pipeline build, and solutions for frequently occurring failures. As with pipeline information, we store this domain knowledge as a class-level attribute in the Knowledge utility class.

\textbf{Historical records of related failed builds (HR)}: These historical records include root causes, the console-log lines from the most downstream failed job that indicate those root causes, and the corresponding solutions. This serves as a few-shot prompting method, where we provide the LLM with examples of the type of answers we expect~\cite{brown_2020_fewShotLearning}. Because different failures have different sets of related historical failed builds, we cannot store this domain knowledge in the Knowledge utility class as we do with other types. Instead, we retrieve records from our History database that share the same most downstream failed job. Then, we use retrieval-augmented generation (RAG)~\cite{rag_Lewis_2020} to rank the top three historical records whose console logs most closely match the current failed build’s logs and provide these records to the LLM.

In the remainder of this section, we describe our experimental design for addressing the research questions.

% -------------------- Subsection ------------------------------
\subsection{Dataset and Methodology for RQ1} \label{subsec_dataset}
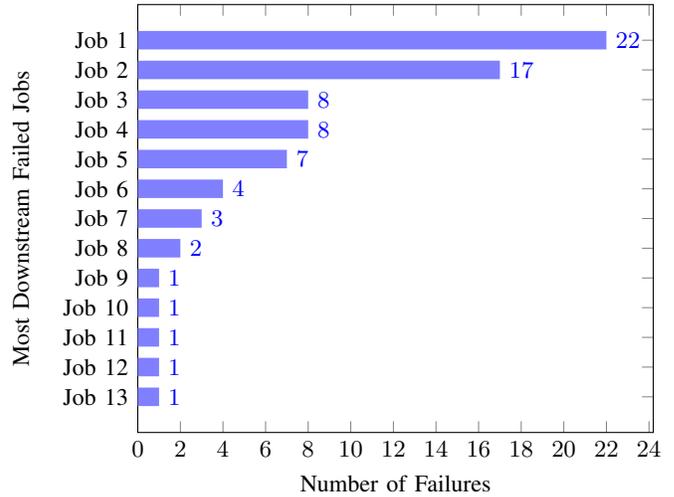
\begin{figure}[htbp]
\centering
\begin{tikzpicture}
\begin{axis}[
    xbar,
    y dir=reverse,
    ylabel style={yshift=6pt},
    xlabel={Number of Failures},
    ylabel={Most Downstream Failed Jobs},
    symbolic y coords={{Job 1},{Job 2},{Job 3},{Job 4},{Job 5},{Job 6},{Job 7},{Job 8},{Job 9},{Job 10},{Job 11},{Job 12},{Job 13}},
    ytick=data,
    xmin=0,
    nodes near coords,
    bar width=7pt,
    tick label style={font=\small},
    label style={font=\small},
    nodes near coords style={font=\small},
    tick align=inside,
    xtick distance=2,
    every axis plot post/.append style={
        draw=none,
        fill=blue!50
    },
]

\addplot coordinates {
    (22,{Job 1}) (17,{Job 2}) (8,{Job 3}) (8,{Job 4}) (7,{Job 5}) (4,{Job 6}) (3,{Job 7}) (2,{Job 8}) (1,{Job 9}) (1,{Job 10}) (1,{Job 11}) (1,{Job 12}) (1,{Job 13})
};
\end{axis}
\end{tikzpicture}
\caption{Frequency distribution of most downstream failed jobs in the collected previous failures.}
\label{fig_datasetFrequencyDistribution}
\end{figure}

We collect all failures in the SAP HANA CI/CD pipeline over a six-month period, yielding $76$ failures associated with $13$ distinct most downstream failed jobs. The frequency distribution of the most downstream failed jobs is shown in \Cref{fig_datasetFrequencyDistribution}. Some jobs triggered failures more frequently. Job 1 accounted for $22$ of the $76$ failures. In contrast, Job 9-Job 13 each occurred only once. The LLM-based failure-management system is tested on its ability to resolve these failures in Jenkins, mirroring the real-world use case. On a side note, these failures occur in the CI/CD pipeline for delivering SAP HANA and do not affect the quality of SAP HANA or any other products provided by SAP. In fact, every product provided by SAP must pass rigorous checks before reaching customers.

For each failure, we interviewed the on-call engineer who resolved it to understand the root cause and how they resolved it. Additionally, we asked them to provide the log lines in the most downstream failed job that indicate the root cause. The root causes, solutions, and log lines indicating the root cause are stored in a Previous Failures document, which we used to evaluate the correctness of root causes and solutions provided by the LLM-based failure-management system.

To automatically provide the LLM with the domain knowledge labeled \enquote{Historical records of related failed builds}, we added each unique failure case to the History database. If a failure occurs multiple times with the same root cause, solution, and log lines indicating the root cause, only the first occurrence is recorded in the database.

In addition, to answer RQ1, we analyzed the Previous Failure document to identify the most frequent root causes. We then asked on-call engineers whether those causes could be prevented from recurring to improve the robustness of the CI/CD pipeline for delivering SAP HANA.

% -------------------- Subsection ------------------------------
\subsection{Methodology for RQ2}
As mentioned earlier, the root causes of failures typically appear in the log of the most downstream failed job. Currently, we use regular expressions~\cite{python_re} to search the console logs for lines that typically contain information about the failed downstream sub-pipeline, remote pipeline, or job (see \Cref{subsec_mostDownstreamJobFinding}). However, this approach will not work if the log format changes. In fact, it has already changed multiple times for the HANA delivery pipeline. Therefore, we explore the capability of an LLM to autonomously identify the most downstream failed job. If successful, it could serve as an alternative when the current regular-expression approach fails.

\begin{figure}[htbp]
\centering
\begin{tikzpicture}[node distance=0.75cm, font=\small]
\node (start) [startstop, xshift=-2cm, text width=5cm] {Find the most downstream failed job.};
\node (llm1) [llm, below of=start, yshift=-0.5cm] {Coordinator \\ Function-calling LLM};

\node (dec1) [decision, below of=llm1, text width=2cm, yshift=-1.75cm] {Is the most downstream failed job reached?};
\node (stop) [startstop, below of=dec1, yshift=-1.75cm, text width=5cm] {Return most downstream failed job};

\node (pro1b) [process, right of=llm1, xshift=3.5cm, text width=3cm] {Retrieve console logs};
\node (llm2) [llm, below of=pro1b, yshift=-0.2cm,] {LLM};
\node (pro3b) [process, below of=llm2, yshift=-0.3cm, text width=3cm] {Find downstream failed job};
% Draw the dashed box
\begin{scope}[on background layer]
\node[dashedbox, fit=(pro1b) (llm2) (pro3b), inner sep=3pt] (group) {};
\node[anchor=center, align=center, inner sep=2pt, fill=white, text width=4cm] (boxtitle)
at ([yshift=10pt]group.north) {LLM-based\\Downstream Job Finder};
\end{scope}

\draw [arrow] (start) -- (llm1);
\draw [arrow] (llm1) -- (dec1);
\draw [arrow] (dec1) -- node[anchor=east] {yes} (stop);
\draw [arrow] (dec1) -- node[anchor=south] {no} (group);
\draw [arrow] (pro1b) -- (llm2);
\draw [arrow] (llm2) -- (pro3b);
\draw [arrow] (group) -- (llm1);
\end{tikzpicture}
\caption{Control-flow diagram of LLM-based most-downstream-failed-job-finding system.}
\label{fig_downstreamJobFindingSystem}
\end{figure}
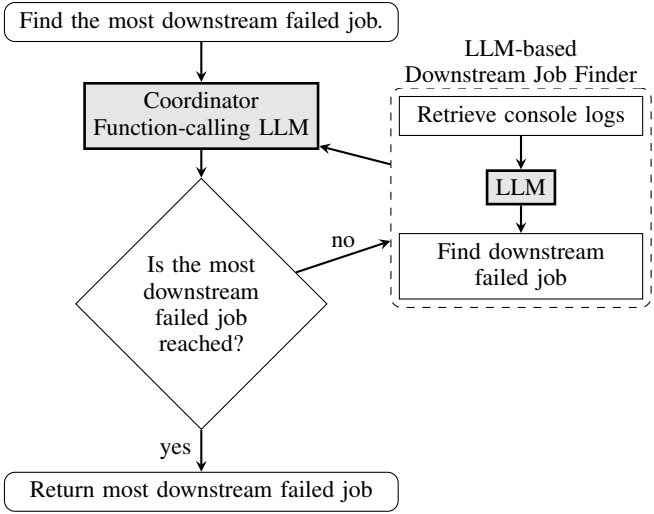

Our LLM-based most-downstream-failed-job-finding system consists of two LLM calls: a coordinating call and a downstream-job-finding call, referred to as Coordinator and Downstream Job Finder. Downstream Job Finder is a subsystem that retrieves a console log and prompts an LLM to identify the failed downstream job within it. Coordinator is an LLM with function-calling capability~\cite{wang_2024}, which allows it to invoke Downstream Job Finder automatically and iteratively until the most downstream failed job is found. The system control-flow diagram is shown in \Cref{fig_downstreamJobFindingSystem}.

In detail, the system works as follows. The Coordinator calls the Downstream Job Finder to locate the failed downstream job, sub-pipeline, or remote pipeline within the CI/CD pipeline. The Downstream Job Finder retrieves the CI/CD pipeline’s console logs, identifies the failed downstream job, sub-pipeline, or remote pipeline (denoted as DS-1), and returns it to the Coordinator. If DS-1 is not the most downstream failed job, the Coordinator calls the Downstream Job Finder to locate the failed downstream job, sub-pipeline, or remote pipeline downstream of DS-1 (denoted as DS-2). This process repeats until it reaches DS-n, the most downstream failed job, at which point the process completes.

To support the LLM-based Coordinator and Downstream Job Finder, in addition to the query, we provide them with instructions. The instruction for Coordinator is shown in \Cref{coordinator_instruction} and the instruction for Downstream Job Finder is shown in \Cref{downstreamJobFinder_instruction}.
% [float, floatplacement=H, caption={Instruction for Coordinator}, label={coordinator_instruction}]
\begin{figure}[htbp]
\center
\begin{lstlisting}
Find the most downstream failed job by iteratively finding downstream jobs/ sub-pipelines/ remote pipelines and following their downstream chains until you reach the most downstream failed job.
\end{lstlisting}
\caption{Instruction for Coordinator}
\label{coordinator_instruction}
\end{figure}

% [float, floatplacement=H, caption={Instruction for Downstream Job Finder}, label={downstreamJobFinder_instruction}]
\begin{figure}[htbp]
\center
\begin{lstlisting}
Your task is to find the failed downstream job or sub-pipeline that caused the pipeline failure and report its name and build number. If you cannot find any downstream job (because the current job is already the most downstream failed job), please return: No failed downstream job found - the job is already the most downstream failed job.
\end{lstlisting}
\caption{Instruction for Downstream Job Finder}
\label{downstreamJobFinder_instruction}
\end{figure}

We test the LLM-based most-downstream-failed-job-finding system using previous failures stored in Jenkins (see \Cref{subsec_dataset} for dataset information) and measure its accuracy rate of identifying the most downstream failed job. Additionally, we evaluate whether the system’s accuracy rate improves when the LLM is provided with domain knowledge, such as pipeline structure, sub-pipeline names, naming conventions, and the log-line format that contains downstream-job information. This domain knowledge is stored as an attribute in the Knowledge utility class, which we access whenever we provide it to the LLM.

% -------------------- Subsection ------------------------------
\subsection{Methodology for RQ3} \label{subsec_rq3method}
There are multiple types of domain knowledge used to manage our CI/CD pipeline failures, including pipeline information, failure-management instructions, and historical records of related failed builds (see \Cref{subsec_method_causeAndSolutionFinding} for a description of these domain-knowledge types). We identify the type of domain knowledge that contributes most to the accuracy of the LLM-based failure-management system. To do so, we apply an ablation study, a method that removes or modifies provided inputs to observe changes in system behavior or performance~\cite{Sheikholeslami_2021}.

We call the LLM-based failure-management system to resolve our previously failed builds stored in Jenkins (see \Cref{subsec_dataset} for dataset information) when given no domain knowledge, one type, two types, and all three types. We then categorize the system's proposed solutions into 3 levels of preferences:
\begin{enumerate}
    \item Red - Incorrect Solution: The proposed solution is incorrect. It either omits the actions needed to solve the problem or recommends incorrect actions that would create additional work, e.g., suggesting incorrect code modifications that would need to be corrected.
    \item Yellow - Indirect solution: The proposed solution resolves the problem, although indirectly. It may require extra effort, but it would not make incorrect modifications that would require manual correction later. For example, a solution that contains the correct fix but also includes an unnecessary step, such as checking access to a service, would be categorized as yellow.
    \item Green - Exact Solution: The proposed solution specifies the exact actions needed to resolve the problem without any unnecessary effort.
\end{enumerate}

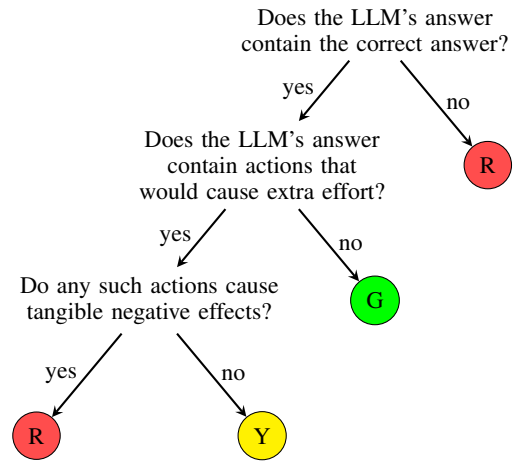
\begin{figure}[htbp]
\centering
\begin{tikzpicture}[node distance=1.5cm, font=\small]
\node (ques1) [question, text width=5cm, align=center] {Does the LLM’s answer contain the correct answer?};
\node (cat1) [categorization, below of=ques1, xshift=1.5cm, yshift=-0.3cm, fill=red!70] {R};
\node (ques3) [question, below of=ques1, xshift=-1.5cm, yshift=-0.3cm, text width=4cm] {Does the LLM’s answer contain actions that would cause extra effort?};
\node (ques4) [question, below of=ques3, xshift=-1.5cm, yshift=-0.3cm, text width=4cm] {Do any such actions cause tangible negative effects?};
\node (cat3) [categorization, below of=ques3, xshift=1.5cm, yshift=-0.3cm, fill=green] {G};
\node (cat4) [categorization, below of=ques4, xshift=-1.5cm, yshift=-0.3cm, fill=red!70] {R};
\node (cat5) [categorization, below of=ques4, xshift=1.5cm, yshift=-0.3cm, fill=yellow] {Y};
\draw [arrow] (ques1) -- node[anchor=east] {yes} (ques3);
\draw [arrow] (ques1) -- node[anchor=west] {no} (cat1);
\draw [arrow] (ques3) -- node[anchor=east] {yes} (ques4);
\draw [arrow] (ques3) -- node[anchor=west] {no} (cat3);
\draw [arrow] (ques4) -- node[anchor=east] {yes} (cat4);
\draw [arrow] (ques4) -- node[anchor=west] {no} (cat5);

\end{tikzpicture}
\caption{Decision tree for evaluating the solution proposed by the LLM.}
\label{fig_decisionTree}
\end{figure}

We categorize the LLM’s responses into three levels using the decision tree shown in \Cref{fig_decisionTree}.

% ----------------------------------------------------------------------------
% Section
% ----------------------------------------------------------------------------
\section{Experiment Results} \label{sec_experimentResult}

% -------------------- Subsection ------------------------------
\subsection{RQ1: What are the most common causes of failure in the CI/CD pipeline for the delivery of SAP HANA?}
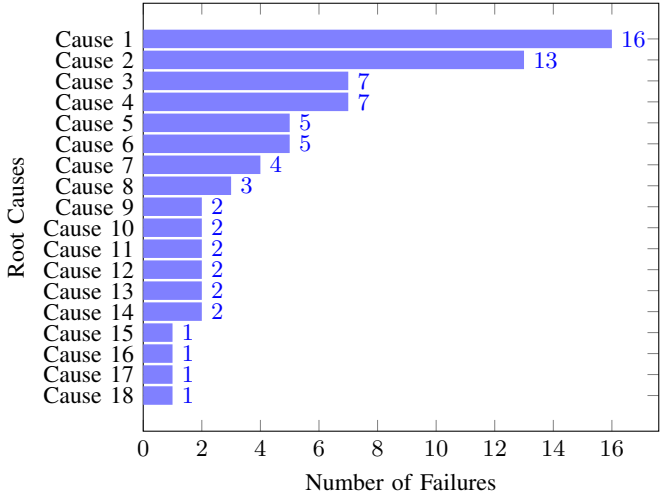
\begin{figure}[htbp]
\centering
\begin{tikzpicture}
\begin{axis}[
    xbar,
    y dir=reverse,
    ylabel style={yshift=0.5pt},
    xlabel={Number of Failures},
    ylabel={Root Causes},
    symbolic y coords={{Cause 1},{Cause 2},{Cause 3},{Cause 4},{Cause 5},{Cause 6},{Cause 7},{Cause 8},{Cause 9},{Cause 10},{Cause 11},{Cause 12},{Cause 13},{Cause 14},{Cause 15},{Cause 16},{Cause 17},{Cause 18}},
    ytick=data,
    xmin=0,
    nodes near coords,
    bar width=7pt,
    tick label style={font=\small},
    label style={font=\small},
    nodes near coords style={font=\small},
    tick align=inside,
    xtick distance=2,
    every axis plot post/.append style={
        draw=none,
        fill=blue!50
    },
]

\addplot coordinates {
    (16,{Cause 1}) (13,{Cause 2}) (7,{Cause 3}) (7,{Cause 4}) (5,{Cause 5}) (5,{Cause 6}) (4,{Cause 7}) (3,{Cause 8}) (2,{Cause 9}) (2,{Cause 10}) (2,{Cause 11}) (2,{Cause 12}) (2,{Cause 13}) (2,{Cause 14}) (1,{Cause 15}) (1,{Cause 16}) (1,{Cause 17}) (1,{Cause 18})
};
\end{axis}
\end{tikzpicture}
\caption{Frequency distribution of causes for the previously collected failures. Cause n, where $1\leq n \leq 18$, denotes the cause of a build failure in the SAP HANA delivery CI/CD pipeline.}
\label{fig_rootCauseFrequencyDistribution}
\end{figure}

Across $76$ previous failures (see \Cref{subsec_dataset}), we identify $18$ distinct causes of failure. The frequency distribution of failure causes is shown in \Cref{fig_rootCauseFrequencyDistribution}.

The most common cause is errors in the Gerrit code-review process for changes that set the container image version for HANA Cloud, causing $16$ of the $76$ failures. A container image is a standalone package that includes everything needed to run this software product on a cloud infrastructure~\cite{xu_2018_containerImage}. Here, a set of tests checks if a container image can be built with the new version. For a Gerrit change to be merged, the tests must pass. If they fail, Gerrit treats the change as not mergeable, and the main CI/CD pipeline fails. The responsibility for this task lies with another team, which is responsible for creating container images. When an error occurs here, the responsible team is contacted to resolve the issue, so that the test finishes successfully. Then, we can restart the CI/CD pipeline.

The second most common cause also occurs frequently, causing $13$ pipeline failures. It occurs in a step that polls a build server to check if the build process for the HANA version to be delivered by the pipeline is complete or still in progress. Once the build is finished, the pipeline continues. In these $13$ cases, however, due to compilation errors during the HANA build, the pipeline step cannot be completed successfully. In such cases, we must contact the responsible development team to resolve the issue with the build, before the CI/CD pipeline can be resumed.

% As in these two most common cases, recurrence cannot be prevented due to the distributed responsibilities within the CI/CD workflow. However, automating the notifications to the responsible teams can increase the overall efficiency of the process.

% -------------------- Subsection ------------------------------
\subsection{RQ2: How accurately can an LLM identify the error location (most downstream failed job) in the CI/CD pipeline for the delivery of SAP HANA?}
In the $76$ previous failures stored in Jenkins (see \Cref{subsec_dataset} for dataset information), $59$ failures are one-round cases, where the most downstream failed job is immediately downstream of the main pipeline (main pipeline $\rightarrow$ most downstream failed job). The remaining $17$ failures are two-round cases, where the most downstream failed job occurs one level further downstream (main pipeline $\rightarrow$ the main pipeline’s failed downstream job $\rightarrow$ most downstream failed job).

Without domain knowledge, the LLM-based most-downstream-failed-job-finding system correctly identified the most downstream failed job in $64$ of $76$ cases. Ten of the twelve failed cases were the fault of the Downstream Job Finder. It had reached the most downstream failed job and could not find further downstream failed job information. However, instead of reporting that it had reached the most downstream failed job, it tried to guess a downstream job from the logs. Then it reported the guessed further downstream job to the Coordinator. The Coordinator tried to access this nonexistent Jenkins job and failed. The remaining two failed cases were the fault of the Coordinator. The Downstream Job Finder reported the most downstream failed job correctly. However, the Coordinator returned the upstream pipeline as the most downstream failed job. We observed no consistent pattern explaining these failures. They occurred in both one-round and two-round cases, and some failed cases had causes and log structures similar to those of successful cases.

When provided with domain knowledge (pipeline structure, sub-pipeline names, naming conventions, and the log-line format that contains downstream job information), the system correctly identified the most downstream failed job in $74$ of $76$ failures. The two failures were both two-round cases. In one case, the Coordinator wrongly returned the upstream pipeline, while the Downstream Job Finder had already provided it with the correct most downstream failed job. In the other case, the Coordinator wrongly stopped after finding a downstream pipeline without calling the Downstream Job Finder again to check whether it had any further downstream jobs. However, these behaviors are inconsistent. After retrying those failed cases, the system successfully found the most downstream failed job.

The variance information of the accuracy rate across different most-downstream failed jobs, in the order (mean, median, interquartile range (IQR), and standard deviation (SD)), is as follows: no domain knowledge: ($0.88, 1.0, 0.17, 0.18$); with domain knowledge: ($0.98, 1.0, 0.0, 0.07$). This shows that the system performed consistently well in both cases, but when provided with domain knowledge, both accuracy and consistency improved. Regarding the time required, the system took a mean of $49.3$ seconds, a median of $47.3$ seconds, an IQR of $11.0$ seconds, and an SD of $9.9$ seconds, indicating that the duration varied only slightly across cases.

% -------------------- Subsection ------------------------------
\subsection{RQ3: Which type of knowledge contributes most to an LLM’s accuracy in proposing solutions for failures in the CI/CD pipeline for the delivery of SAP HANA?}

\begin{figure}[htbp]
\centering
\begin{tikzpicture}
\begin{axis}[
    ybar stacked,
    width=\columnwidth,
    xlabel={Types of Domain Knowledge},
    ylabel={Number of Builds},
    xlabel style={yshift=5pt},
    ytick distance=10,
    symbolic x coords={None, PI, FMI, HR, {PI\\+\\FMI}, {PI\\+\\HR}, {FMI\\+\\HR}, {PI\\+\\FMI+HR}},
    xtick=data,
    xticklabel style={anchor=north, align=center},
    bar width=15pt,
    enlarge x limits=0.1,
    nodes near coords,
    every node near coord/.append style={
        font=\scriptsize,
        color=black,
        /pgf/number format/precision=0,
    },
    nodes near coords align={center},
    legend style={at={(0.5,-0.27)}, anchor=north, legend columns=-1, font=\small},
    legend style={font=\small},
    tick label style={font=\small},
    label style={font=\small},
    nodes near coords style={font=\tiny},
    clip=false,
    ymin=0,
    /pgfplots/stack plots/y expr/.code=\pgfmathparse{round(#1)},
    every axis plot post/.append style={
        fill opacity=1,
        draw=black,
        line width=0.3pt
    }
]

\addplot+[ybar, fill=green] coordinates {(None,13) (PI,15) (FMI,40) (HR,70) ({PI\\+\\FMI},33) ({PI\\+\\HR}, 70) ({FMI\\+\\HR}, 70) ({PI\\+\\FMI+HR}, 69)};
\addplot+[ybar, fill=yellow] coordinates  {(None,13) (PI,21) (FMI,27) (HR,4) ({PI\\+\\FMI},33) ({PI\\+\\HR}, 6) ({FMI\\+\\HR}, 6) ({PI\\+\\FMI+HR}, 5)};
\addplot+[ybar, fill=red!70] coordinates{(None,50) (PI,40) (FMI,9) (HR,2) ({PI\\+\\FMI},10) ({PI\\+\\HR}, 0) ({FMI\\+\\HR}, 0) ({PI\\+\\FMI+HR}, 2)};

\legend{Exact Solution, Indirect Solution, Incorrect Solution}

\end{axis}
\end{tikzpicture}
\caption{Ablation study results. None: no domain knowledge provided; PI: pipeline information provided; FMI: failure-management instructions provided; HR: historical records of related failed builds provided; PI+FMI: PI and FMI provided; PI+HR: PI and HR provided; FMI+HR: FMI and HR provided; PI+FMI+HR: all domain-knowledge types (PI, FMI, and HR) provided.}
\label{fig_ablation_study}
\end{figure}
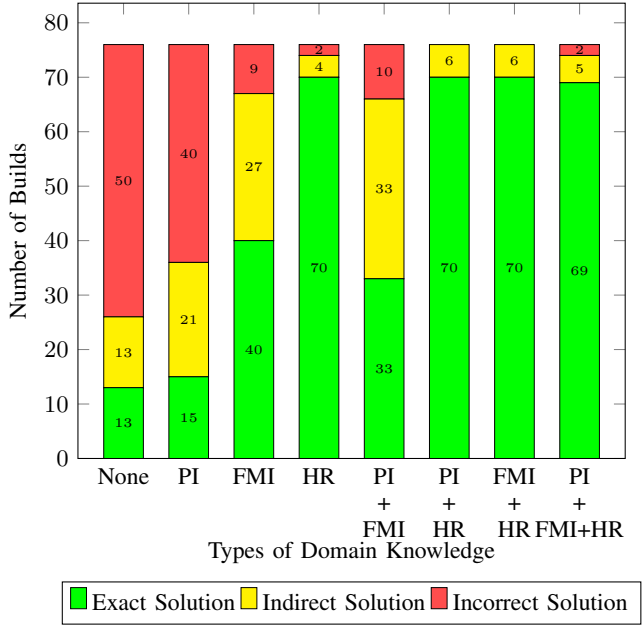

We run the LLM-based failure-management system to identify root causes and propose solutions for $76$ previous failures stored in Jenkins (see \Cref{subsec_dataset}). We evaluate all eight combinations of domain knowledge: no domain knowledge; each single type (pipeline information, failure-management instructions, or historical records); each pair of types; and all three types together. Descriptions of these domain-knowledge types appear in \Cref{subsec_method_causeAndSolutionFinding}. The results are shown in \Cref{fig_ablation_study}.

% On average, the system took $1.19$ minutes to send the root causes and solutions to our on-call engineers. The median duration was $1.16$ minutes, the interquartile range was $0.14$ minutes, and the standard deviation was $0.14$ minutes, indicating that the duration varied only slightly across cases. 

Regarding the time required, the system took a mean of $1.19$ minutes, a median of $1.16$ minutes, an IQR of $0.14$ minutes, and an SD of $0.14$ minutes, indicating that the duration varied only slightly across cases. Regarding the accuracy of the proposed solutions, we observed the worst performance when no domain knowledge was provided: the system produced incorrect solutions for $50$ failures, indirect solutions for $13$ failures, and exact solutions for $13$ failures (see the definitions of exact, indirect, and incorrect solutions in \Cref{subsec_rq3method}).
% These $13$ failures contained clear error messages in the console logs of the most downstream failed jobs, and all ablation conditions produced exact solutions for them.

Among the three single-domain conditions, the HR condition performed best, producing $70$ exact solutions, $4$ indirect solutions, and $2$ incorrect solutions. In the incorrect cases, the system suggested some correct actions but also proposed incorrect code modifications. The PI condition produced $15$ exact, $21$ indirect, and $40$ incorrect solutions - slightly better than the condition without domain knowledge. The FMI condition outperformed the PI condition, producing $40$ exact, $27$ indirect, and $9$ incorrect solutions, likely because FMI contains solutions for frequently occurring failures.

Among the three pairwise conditions, the PI+FMI combination performed worst: $33$ exact, $33$ indirect, and $10$ incorrect solutions - worse than FMI alone. Both the PI+HR and the FMI+HR combinations yielded $70$ exact, $6$ indirect, and $0$ incorrect solutions. The condition with all three domain-knowledge types performed worse than the pairwise conditions that included historical records. It produced $69$ exact, $5$ indirect, and $2$ incorrect solutions, compared with $70$, $6$, and $0$ for PI+HR and FMI+HR, respectively.

To further statistically analyze the variances in accuracy between different root causes, we assigned three levels of correctness a score: exact = $1$, indirect = $0.5$, and incorrect = $0$. For each cause of failure, we calculated its accuracy rate by dividing the total accuracy score by the number of cases failed by that cause. The results are shown in \Cref{table_ablation_statistics}.

\begin{table}[htbp]
\caption{Ablation Condition Accuracy Rate Variation Across Different Root Causes}
\label{table_ablation_statistics}
\centering
\setlength{\tabcolsep}{1.75pt}  % default is 6pt
\begin{tabular}{|c|c|c|c|c|c|c|c|c|}
\hline
\textbf{Metrics} & \textbf{\textit{None}}& \textbf{\textit{PI}}& \textbf{\textit{FMI}}& \textbf{\textit{HR}}& \textbf{\textit{PI+FMI}}& \textbf{\textit{PI+HR}}& \textbf{\textit{FMI+HR}}& \textbf{\textit{PI+FMI+HR}} \\
\hline
Mean& 0.37& 0.48& 0.7& 0.95& 0.65& 0.96& 0.96& 0.88\\
\hline
Median& 0.25& 0.5& 0.77& 1.0& 0.63& 1.0& 1.0& 1.0\\
\hline
IQR& 0.61& 0.69& 0.5& 0.0& 0.5& 0.0& 0.0& 0.16\\
\hline
SD& 0.39& 0.36& 0.33& 0.15& 0.37& 0.12& 0.13& 0.26\\
\hline
\end{tabular}
\label{tab1}
\end{table}

These results show that ablation conditions containing HR - including HR, PI+HR, FMI+HR, and PI+FMI+HR - consistently provide correct answers and exhibit only small variance across different root causes. In contrast, ablation conditions that do not contain HR - including None, PI, FMI, and PI+FMI - show notably lower accuracy and higher variability across different causes of failure. This indicates that the accuracy of ablation conditions without HR strongly depends on the root cause. Root causes with clearer or more familiar error messages achieve significantly higher accuracy rates than others. Overall, historical records of related failed builds contributed most to the LLM’s accuracy in proposing solutions. Every condition that included historical records showed a substantial improvement in accuracy compared with conditions that did not include them.

% ----------------------------------------------------------------------------
% Section
% ----------------------------------------------------------------------------
\section{Lessons Learned} \label{sec_lessonLearned}
% Length: ~ 1 pages
Through this study, we observe that LLMs can support the automated identification of the most downstream failed job that causes a pipeline failure, the extraction of root causes from console logs, and the proposal of precise solutions for failures in the CI/CD pipeline for delivering SAP HANA. Additionally, we believe that with function-calling capabilities, the LLM-based system can execute solutions, enabling end-to-end automated failure management of the CI/CD pipeline. This requires two key components: function-calling capabilities and domain knowledge. First, function calling allows the LLM to retrieve the necessary information for failure management and to execute solutions. Second, our experimental results demonstrate that incorporating domain knowledge enhances the performance of the LLM-based system. Historical records of previous failed builds and their solutions increase the accuracy of LLM-proposed solutions, while pipeline structure, sub-pipeline names, naming conventions, and the log-line format that contains downstream job information improve the system’s ability to identify the most downstream failed job.

Additionally, we observe that even when provided with historical records and asked to follow recorded solutions exactly, the LLM still adds extra steps. In some cases, the steps are intended to prevent failures from recurring or to mitigate negative effects (e.g., adding error handling or logs). These are helpful suggestions, so we categorize these solutions as green. However, in some cases, these steps are extra actions taken before following the exact solutions mentioned in the historical records. Sometimes these extra actions are incorrect and require manual correction. This behavior is also observed in other studies~\cite{roy_2024, chaudhary_2024_LLMChatbot, xu_2025}. Therefore, it should be considered when allowing such an LLM-supported failure-management system to apply the suggested solutions automatically.

We observe that the accuracy of PI+FMI+HR is lower than that of PI+HR and FMI+HR, and the accuracy of PI+FMI is lower than that of FMI alone. This means that adding more domain knowledge does not always increase an LLM’s accuracy. Therefore, it is necessary to evaluate the contribution of each type of domain knowledge to the LLM performance. Moreover, in the CI/CD pipeline for the delivery of SAP HANA, we observe that some failures recur frequently. Solving these failures with an LLM would be cost-inefficient, so we prefer non-LLM methods. Therefore, we plan to construct a hybrid LLM failure-management system as follows. First, the system uses a non-LLM approach to find the most downstream failed job and extract its console log. If the non-LLM approach fails, the system calls the LLM. Once the most downstream failed job is found, the system checks whether a similar failure exists in historical records. If a match is found, it applies the historical solution and observes whether the rebuild succeeds. If no similar historical failure exists, or if the rebuild fails, the system calls the LLM.

\section{Threats to Validity} \label{sec_threadsToValidy}
% Length: 0.5-1 page

\subsection{Construct Validity}
The domain knowledge we provided for the ablation study may not capture the full complexity of our CI/CD workflow. In particular, our setup without the historical failure data potentially yields less accurate results than such an LLM-based solution could generally provide. Hence, we cannot rule out that a more comprehensive description of the pipeline would have led to better results. Furthermore, it should be considered that we conducted the experiments using the GPT-4o model with temperature 0.0 only, and results may vary with other large language models and with different temperatures. Yet, the insights we have gained from these experiments already provide an outlook on what is potentially possible when such an LLM-based approach is enriched with the appropriate domain knowledge.

\subsection{Internal Validity}
Since there are only four to six software releases per week, we were able to collect just a limited amount of data from failed pipelines during the phase of the study. Despite that, we decided to conduct the experiments with this available data because it already allowed us to record different variants of pipeline failures, whose causes occurred at several steps in the main pipeline and in the sub-pipelines. While we believe this study provides valuable insights, it is important to consider that the technical CI/CD landscape for SAP HANA is constantly changing and evolving, which could in future lead to different outcomes with the present setup for the experiments. Because SAP HANA is a large software project, the results we achieved could be of interest to pursue further research into LLM-supported CI/CD pipeline failure management.

\subsection{External Validity}
Our study is not representative of all other areas where automated CI/CD workflows are involved, regardless of whether Jenkins or other tools are used. The setup of our Jenkins pipeline is specifically tailored to the organizational structure of the CI/CD workflow for SAP HANA. Responsibility and control for different parts of our CI/CD workflow are subject to cross-team and cross-functional dependencies. We think that for other organizations facing comparable conditions, our study can provide useful insights to consider a prospective option for automated CI/CD failure handling.

% ----------------------------------------------------------------------------
% Section
% ----------------------------------------------------------------------------
\section{Conclusion} \label{sec_conclusion}
This study tests the potential of large language models (LLMs) to automate failure management in the CI/CD pipeline for the delivery of SAP HANA. First, the experiment results show that the most common causes of failures are errors triggered by remote services. Therefore, we can not fully prevent recurrence. Second, an LLM with function-calling capabilities~\cite{wang_2024} can identify the most downstream failed job correctly with small variance across different most-downstream failed jobs (mean, median, interquartile range (IQR), standard deviation (SD)) = ($0.88, 1.0, 0.17, 0.18$) without domain knowledge, and performs even better with domain knowledge: ($0.98, 1.0, 0.0, 0.07$). Third, historical records of related failed builds contribute most to the accuracy of the LLM’s proposed solutions. Ablation conditions that involve historical records produce higher accuracy with lower variance across different root causes compared to ablation conditions without historical records. The worst-performing ablation condition that involves historical records is (mean, median, IQR, SD) = ($0.88, 1.0, 0.16, 0.26$), while the best-performing ablation condition that does not involve historical records is ($0.7, 0.77, 0.5, 0.33$).

% Second, an LLM with function-calling capabilities~\cite{wang_2024} identifies the most downstream failed job with $97.37\%$ accuracy when provided with domain knowledge and $84.21\%$ accuracy when not provided with domain knowledge. Third, historical records of related failed builds contribute most to the accuracy of the LLM's proposed solutions. In the highest-accuracy ablation condition that included historical records, the system produces $70$ exact and $6$ indirect solutions out of $76$ recorded failures. By contrast, the highest-accuracy condition that does not include historical records produces only $40$ exact, $27$ indirect, and $9$ incorrect solutions.

We observe that an LLM-based failure-management system can automate handling failures in the CI/CD pipeline for delivering SAP HANA. However, because some failures recur frequently, it is more cost-effective to use a hybrid approach that first attempts non-LLM solutions and invokes the LLM only if those approaches fail. We encourage future studies using larger datasets and incorporating more domain knowledge, in combination with different LLMs and temperatures, to validate and generalize these findings. Moreover, we encourage research into hybrid LLM failure-management systems that operate cost-effectively.

\newpage

\bibliographystyle{IEEEtran}
\bibliography{references}

\end{document}